\documentclass[11pt]{article}
\usepackage{jheppub}
\usepackage[utf8]{inputenc}
\usepackage{amsmath}
\usepackage{graphicx}
\usepackage{xcolor}
\usepackage{comment}
\usepackage[toc,page]{appendix}
\usepackage{feynmp}
\usepackage{tikz-feynman}
\tikzfeynmanset{compat=1.1.0}
\usetikzlibrary{positioning,arrows}
\usetikzlibrary{decorations.pathmorphing}
\usetikzlibrary{decorations.markings}

\graphicspath{{Images/}}
\title{AdS Higgs mechanism from double trace deformed CFT}
\date{\today}
\author{Andreas Karch,}
\author{Mianqi Wang and}
\author{Merna Youssef}
\affiliation{Theory Group, Weinberg Institute, Department of Physics, University of Texas,\\
\phantom{}\hspace{0.5cm}2515 Speedway, Austin, TX 78712, USA.}
\emailAdd{karcha@utexas.edu}
\emailAdd{mqwang@utexas.edu}
\emailAdd{myoussef@utexas.edu}
\abstract
{Explicit breaking of a global symmetry in a conformal field theory is holographically dual to giving mass to a gauge field living in AdS via the Higgs mechanism. We show that if this breaking is induced via a double trace deformation the Higgs mechanism is induced via a scalar loop diagram. The mass can be calculated analytically in both bulk and field theory and we find perfect agreement. While representing familiar physics, the mechanism is  identical to how the graviton picks up a mass in the holographic dual of a conformal field theory coupled to a bath.
}

\begin{document}
\tikzset{
line/.style={thick, decorate, draw=black,}
 }
\maketitle
\section{Introduction}
Gauge invariance is the main concept underlying the standard model which describes the theory of electroweak and strong interactions. Typically gauge invariance implies the existence of massless spin 1 vector bosons. The Higgs mechanism is a phenomenon by which such a spin 1 gauge field acquires mass upon spontaneous breaking of the gauge symmetry. The massless vector field joins with a massless scalar to form a massive vector field. In the limit when the gauge coupling goes to zero, the longitudinal degrees of freedom of what was the massive spin 1 field becomes the Goldstone boson of what now is a spontaneously broken global symmetry, leaving the vector field massless.

The AdS/CFT correspondence postulates an equivalence between a $d$-dimensional conformal field theory (CFT) and gravity on AdS$_{d+1}$. Global symmetries in the CFT imply existence of massless gauge fields in the bulk. The global symmetry of the CFT is dual to a gauge symmetry in the bulk, with gauge transformations that do not vanish at the AdS boundary playing the role of genuine global symmetries; all others being, as usual, simply a redundancy of the description. The operator/field map of AdS/CFT establishes a duality between the gauge field in the bulk and a global symmetry current on the boundary. Masslessness of the former is dual to conservation of the latter, also fixing the scale dimension of the current to be $d-1$.

To study breaking of the global symmetry in the CFT we also need to consider a charged operator $O$. If we add $O$ to the Lagrangian,
\begin{equation}
\label{tree}
{\cal L} = {\cal L}_0 - h O +c.c.    
\end{equation}
the symmetry is explicitly broken. 
The coupling $h$ is chosen to be real.

If instead of adding $O$ to the Lagrangian, $O$ acquires an expectation value, then the symmetry is spontaneously broken. While physically very distinct, in the bulk both are dual to the standard Higgs mechanism induced by the charged scalar $\phi$ dual to $O$. $\phi$ develops a non-vanishing, position dependent expectation value. The difference between explicit and spontaneous breaking is encoded in the near boundary behavior of $\phi$: for explicit breaking the non-vanishing profile has the leading, non-normalizable term turned on. For spontaneous breaking this leading term vanishes. In either case, the bulk vector becomes massive due to the Higgs mechanism triggered by $\phi$.

In this work, we instead study explicit breaking of the global symmetry by a double trace deformation. We introduce two charged operators $O_1$ and $O_2$ of charge $q_1$ and $q_2$ respectively. Both $O_1$ and $O_2$ are single-trace operators, that is they are made gauge invariant under the $SU(N)$ gauge symmetry underlying the CFT in AdS/CFT via a single trace. Such single trace operators are the basic building blocks in the large $N$ limit. The Lagrangian we study is
\begin{equation}
\label{singlesided}
{\cal L} = {\cal L}_0 - h O_1 O_2 + c.c.    
\end{equation}
For simplicity, we focus on the case of a marginal double trace, that is the dimensions $\Delta_1$ and $\Delta_2$ of $O_1$ and $O_2$ respectively obey $\Delta_1+\Delta_2=d$. As we will see, this time the dual scalars $\phi_1$ and $\phi_2$ do not get any tree level expectation values. The gauge field will nevertheless get a mass via the Higgs mechanism, which in this case is only induced by a one loop effect -- while $\phi_1$ and $\phi_2$ by themselves have zero expectation value, the bi-linear condensate $\langle \phi_1 \phi_2 \rangle$ is non vanishing. For fermionic matter this calculation in the bulk has first appeared in \cite{Rattazzi:2009ux}. Besides studying this for the simpler case of the scalar we will also give an independent calculation on the field theory side yielding perfect agreement, giving a reassuring check of the AdS/CFT framework.

A closely related scenario one can study is that of two CFTs coupled via a charged double trace deformation. One can think of one CFT as the CFT of interest, the other as a bath. The Lagrangian in this case reads
\begin{equation}
\label{doublesided}
{\cal L} = {\cal L}_1 + {\cal L}_2 - h O_1 O_2  + cc  . \end{equation}
Here ${\cal L}_i$ with $i=1,2$ are two decoupled CFTs and this time the operators $O_i$ are only made out of the fields of the $i$-th CFT. Each CFT has a separate conserved $U(1)$ symmetry. Upon coupling the two CFTs via the double trace the global symmetry gets broken from $U(1) \times U(1)$ to a single $U(1)$ under which $O_1 O_2$ is neutral. In the holographically dual picture we start with two separate AdS spaces coupled to each other only via boundary conditions. We will refer to this scenario as the double-sided scenario, and the previous one from \eqref{singlesided} as the single-sided scenario.

This latter scenario of a CFT coupled to a bath has received a lot of attention recently. It is the arena in which non-trivial saddles of the Euclidean path integral responsible for the unitary evaporation of black holes, the so called entanglement islands, can be found \cite{Penington:2019npb,Almheiri:2019psf}. It is also the holographic dual for KR-branes \cite{Karch:2000ct,Karch:2000gx}, which in turn are a testbed for the study of entanglement islands, in particular in higher dimensions. In this case,  not just a bulk vector boson but even the graviton itself gets a mass via a similar loop induced Higgs mechanism \cite{Porrati:2001db,Porrati:2003sa}. In the field theory, the decoupled system has two independent diffeomorphism invariances which get broken down to the diagonal diffeomorphism invariance, similar to the case of the $U(1) \times U(1) \rightarrow U(1)$ breaking of the vector: the graviton dual to the stress tensor in the original CFT gets a mass. The mass in this case can be analytically calculated in both field theory and bulk and exact agreement has been found in \cite{Aharony:2006hz}. By repeating this calculation for the case of the standard Higgs mechanism dual to the explicit breaking of a global symmetry we hope to somewhat demystify this mechanism.

This paper is organized as follows: we begin in section \ref{sec:bg} by reviewing some background material on AdS boundary conditions and Noether's theorem, which will prove helpful in the computations of the following section. In section \ref{sec:cft}, we discuss the field theory computations of the mass induced by double trace deformation in the single- and the double-sided setup. In section \ref{sec:bulk}, we compute the mass on the gravity side, and find a perfect match with the field theory results.

\section{Background}\label{sec:bg}
\subsection{Boundary Conditions in AdS}
An important role in our calculation will be played by boundary conditions on fields in AdS. For a scalar field dual to an operator of dimension $\Delta_i$ the asymptotic behavior near the boundary of AdS goes like 
\begin{equation}
\label{asymp}
    \phi_i(z\rightarrow 0)\sim \alpha_i(x)z^{d-\Delta_i}+\beta_i(x)z^{\Delta_i}
\end{equation} 
in Poincare patch coordinates
\begin{equation}
ds^2=z^{-2}(dz^2+dx^\mu dx_\mu).
\end{equation}
Focusing on the case $\Delta_i > d/2$ the second term in \eqref{asymp} vanishes faster and is ``normalizable" with the standard norm, that is it leads to finite energy fluctuations. 
The first term on the other hand, though also tending towards zero, is non-normalizable.
The case of $d/2-1<\Delta_i <d/2$ corresponds to what is known as the alternative quantization \cite{Klebanov:1999tb}. In this case, the role of the two terms swapped in that this time the first term vanishes faster. Nevertheless, in this case one can use an alternative norm which still leaves the second term to be the normalizable one. The expansion \eqref{asymp} with $\alpha$ being the non-normalizable term is valid for the entire allowed range of $\Delta \geq d/2-1$.

In the absence of sources in the field theory we simply demand that the non-normalizable mode has to die out, in other words, the undeformed boundary condition for those fields is $\alpha_i(x)=0$. Note that this boundary conditions is consistent with $\phi_i(z) \equiv 0$ -- the scalar field vanishes identically. If, as in the case considered in this paper, $O_i$ and hence $\phi_i$ are charged under a global symmetry, an identically vanishing scalar is the only configuration that leaves the gauge symmetry in the bulk unbroken.

If the global symmetry is explicitely broken by a single trace operator as in \eqref{tree}, we still have a Dirichlet boundary condition which this time reads $\alpha(x)=h$. To find the actual profile $\phi(z)$ we need to solve the equations of motion for $\phi(z)$ given this boundary condition. The details of the setup, in particular the potential $V(\phi)$, become important and differ from example to example. But one thing is for certain: the identically vanishing solution $\phi \equiv 0$ is no longer allowed. The gauge symmetry in the bulk is broken via the Higgs mechanism and the vector gauge field picks up a mass at tree level.

If we instead add a double trace deformation as in either the single-sided scenario of \eqref{singlesided} or the double-sided one of \eqref{doublesided}, the boundary conditions become
\cite{Aharony:2001pa,Witten:2001ua}
\begin{equation}
\label{bc}
    \alpha_1(x)=h(2\Delta_2-d)\beta^*_2(x),\qquad \alpha_2(x)=h(2\Delta_1-d)\beta^*_1(x).
\end{equation}
This boundary conditions where originally derived for a canonically normalized scalar field $X$ with kinetic term $(\partial X)^2/2$. The equations \eqref{bc} for a complex scalar follow for a canonically normalized complex scalar by writing it out as two real scalars
\begin{equation}
\phi = \frac{X + iY}{\sqrt{2}},
\end{equation}
wich kinetic terms
\begin{equation}
    \label{complexaction}
    {\cal L} = -\frac{1}{2} \left[ (\partial X)^2 + (\partial Y)^2 \right ] = - (\partial \phi) (\partial \phi^*) 
\end{equation}
and similarly spelling out the full double trace deformation in terms of real operators 
$O_i = (O_{Xi} + O_{Yi})/\sqrt{2}$
\begin{equation}
h (O_1 O_2 + O_1^* O_2^*)
= h \left ( O_{X1} O_{X2}  - O_{Y1} O_{Y2} \right ). 
\end{equation}
In the double-sided case thes boundary conditions \eqref{bc} glue one AdS to the other; charge and energy can flow from one AdS to the other. These boundary conditions are often referred to as ``transparent". In the single sided case  \eqref{bc} simply relates the near boundary behavior of the two fields living in a single AdS space.

Note that, unlike in the case of the breaking by a single trace operator, the boundary conditions \eqref{bc} still allow the identically vanishing solution $\phi \equiv 0$ -- the vector field remains massless at tree level.
Naturally the question arises here how one would generate mass for the vector fields in the bulk from the boundary condition \eqref{bc}. Following Porrati's calculation in the case of a graviton \cite{Porrati:2003sa}, we will show that the vector will in fact get a mass via a 1-loop diagram; one can think of this as a bi-linear condensate of the scalar field that is non-vanishing even when the tree-level expectation value $\langle \phi_i \rangle$ vanishes identically. The relevant Feynman diagrams for the single- and double-sided case are depicted panel a and b of Figure 1 respectively.
  
The fact that the double trace deformation is sub-leading in the large $N$ limit underlying AdS/CFT allows one to calculate the vector mass analytically order by order in the coupling constant $h$. In the bulk this simply amounts to calculating the 1-loop correction to the vector boson mass. On the boundary side, we can calculate the dual quantity, the anomalous dimension of the no longer conserved current, perturbatively in the coupling $h$. For the case of the massive graviton and its holographically dual non-conserved stress tensor this was done before in \cite{Aharony:2006hz}. Our calculation proceeds virtually identical to the one in \cite{Aharony:2006hz}, even though the fact that we have to deal with much fewer tensor structures leads to significant technical simplifications.

In the calculation of \cite{Aharony:2006hz}, the massless graviton before Higgsing is dual to a conserved spin 2 operator (the stress tensor) $\partial_\mu T_{\mu\nu} = 0$. When the double trace deformation of the double sided case \eqref{doublesided} breaks one of the diffeomorphism symmetries, we get that one of the two originally conserved stress tensors is no longer conserved, $\partial_\mu T_{\mu \nu} = K_\nu$ where $K_\nu$ is a vector operator dual to a vector field in the bulk that joins with the graviton dual to $T_{\mu \nu}$ to form a massive graviton. 
In the same manner, one photon obtains a mass under the same deformation in both double- and single-sided case. Before the deformation, we are guaranteed to have a conserved current $\partial_\mu J^\mu = 0$. This current operator on the boundary maps to the massless vector field in the bulk and is expected to be of dimensions $d-1$. However, after deformation, we obtain a non-conserved current.

\begin{center}
\begin{tikzpicture}[captiontext/.style={below=3mm, text width=5cm}]\label{fig:loop}
  \begin{feynman}
    \vertex (x);
    \vertex[right=1.5cm of x] (y);
    \vertex[ left=of x] (a);
    \vertex[below left=of x] (b);
    \vertex[above right=of y] (c);
    \vertex[right=of y] (d);
    \vertex[above right= 0.5cm and 0.75cm of x] (p) ;
    \vertex[below right= 0.5cm and 0.75cm of x] (p1); 
   \node[above, blue] at (p) {$\phi_1$};
\diagram*{
        (x) --[scalar, bend left] (p),
        (y) --[scalar, bend right] (p),
        (x) --[scalar, half right] (y),
        (x) --[photon] (a),
        (y) --[photon] (d),
    };
    \node[below, blue] at (p1) {$\phi_2$};
  \end{feynman}
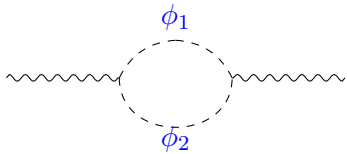
  \node[captiontext] at (current bounding box.south) {Figure 1a: the single-sided scenario: one AdS where the two scalar fields live.};
\begin{scope}[xshift=7cm, local bounding box=g]
  \begin{feynman}
    \vertex (x);
    \vertex[right=1.5cm of x] (y);
    \vertex[ left=of x] (a);
    \vertex[below left=of x] (b);
    \vertex[above right=of y] (c);
    \vertex[right=of y] (d);
    \vertex[above right= 0.5cm and 0.75cm of x] (p) ;
    \vertex[below right= 0.5cm and 0.75cm of x] (p1); 
  \node[above left, blue] at (p) {$\phi_1$};
\diagram*{
        (x) --[scalar, bend left] (p),
        (y) --[scalar, bend right] (p),
        (x) --[scalar, half right] (y),
        (x) --[photon] (a),
        (y) --[photon] (d),
    };
    \node[below left, blue] at (p1) {$\phi_1$};
    \node[below right, blue] at (p1) {$\phi_2$};
    \node[above right, blue] at (p) {$\phi_2$};
  \end{feynman}
  \draw [solid] (.75,-1) -- (.75,1);
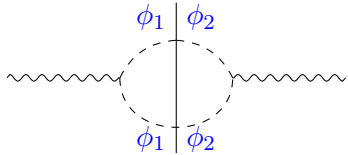
  \node [captiontext] at (g.south) {Figure 1b: the double-sided scenario: two copies of AdS where fields can move from one AdS to the other.};
  \end{scope}
\end{tikzpicture}
\end{center}

\subsection{Noether's theorem}

Noether theorem is a statement about the invariance of the action. The easiest way for the action to be invariant is for the Lagrangian itself to be invariant. In general it is sufficient if the Lagrangian transforms by a total derivative, but we will not need to consider this possibility for our application. Symmetries of the Lagrangian ensure a conserved quantity. 

In our application we will be interested in a variant of Noether's theorem that quantifies the non-conservation of the current once the symmetry is explicitly broken, but to establish notation let us briefly review the textbook case of the conserved current that results from an unbroken symmetry of the Lagrangian.
Let's start with a Lagrangian defined in terms of a field $\phi$ and its derivatives $\partial_\mu \phi$. The symmetry equation is $\delta_\alpha  \mathcal{L} =0 $ which explicitly can be written as:
\begin{equation}
\begin{aligned}\\&
    \frac{\delta \mathcal{L}}{\delta (\partial_\mu \phi)}\delta_\alpha(\partial_\mu \phi)+\frac{\delta \mathcal{L}}{\delta ( \phi)} \delta_\alpha(\phi) = 0 
\label{symm}
\end{aligned}
\end{equation}
where
\begin{equation}
    \delta_\alpha (\partial_\mu \phi) = \partial_\mu (\delta_\alpha \phi) .
\end{equation}
Now if we apply the chain rule we can rewrite this as
\begin{equation}
\partial_\mu(\frac{\delta \mathcal{L}}{\delta ( \partial_\mu\phi)}\delta_\alpha\phi) - \partial_\mu(\frac{\delta \mathcal{L}}{\delta (\partial_\mu\phi)})\delta_\alpha\phi+\frac{\delta \mathcal{L}}{\delta ( \phi)} \delta_\alpha(\phi) = 0
\end{equation}
and the equation of motion (EOM) will take the form 
\begin{equation}
    \partial_\mu \frac{\delta \mathcal{L}}{\delta(\partial_\mu \phi)}- \frac{\delta \mathcal{L}}{\delta\phi}=0 
    \label{eom}
\end{equation}
\begin{equation}
    \partial_\mu (\frac{\delta \mathcal{L}}{\delta(\partial_\mu \phi)}) \delta_\alpha\phi  = \delta_\alpha \phi \frac{\delta \mathcal{L}}{\delta \phi} = -\frac{\delta \mathcal{L}}{\delta (\partial_\mu \phi)}\delta_\alpha(\partial_\mu \phi)
    \label{totder}
\end{equation}
where the first equality is by the EOM and multiplying by $\delta_\alpha \phi$ and the second equality is through the symmetry equation. Taking the first and last term we get:
\begin{equation}
   \partial_\mu (\frac{\delta \mathcal{L}}{\delta(\partial_\mu \phi)}) \delta_\alpha\phi+ \frac{\delta \mathcal{L}}{\delta (\partial_\mu \phi)}\delta_\alpha(\partial_\mu \phi) = 0 = \partial_\mu (\frac{\delta \mathcal{L}}{\delta(\partial_\mu \phi)}\delta_\alpha\phi)
   \label{cons}
\end{equation}
or in other words, the current 
\begin{equation}
J^{\mu} = \frac{\delta \mathcal{L}}{\delta(\partial_\mu \phi)}
\end{equation}
is conserved.

Following the same steps in the presence of an explicit symmetry breaking term in the Lagrangian will gives us a non-trivial right hands side in the current conservation equation $\partial_{\mu} J^{\mu} =0$ in terms of the variation of the symmetry breaking terms.

\subsection{Deformed Noether theorem}

What happens if we add a symmetry breaking deformation to the Lagrangian? In our application below we will be interested in the double trace deformed Lagrangians \eqref{singlesided} and \eqref{doublesided}, but to explain the basic principle let us first consider the single trace deformation of \eqref{tree} with a single symmetry violating operator  $O$.  The variation of the Lagrangian no longer vanishes, but instead is given by
\begin{equation}
    \delta_\alpha \mathcal{L} = \frac{\delta\mathcal{L}}{\delta O} \, \delta_\alpha O =\kappa 
\end{equation}
So now the left hand side of \eqref{symm} will not be zero; it will be $\kappa$ and therefore the last term in equation \eqref{totder} will pick up a $-\kappa$ term to become:
\begin{equation}
    \partial_\mu (\frac{\delta \mathcal{L}}{\delta(\partial_\mu \phi)}) \delta_\alpha\phi  = \delta_\alpha \phi \frac{\delta \mathcal{L}}{\delta \phi} = -\frac{\delta \mathcal{L}}{\delta (\partial_\mu \phi)}\delta_\alpha(\partial_\mu \phi)-\kappa
\end{equation}
With this equation \eqref{cons} becomes
\begin{equation}
    \partial_\mu (\frac{\delta \mathcal{L}}{\delta(\partial_\mu \phi)}) \delta_\alpha\phi+ \frac{\delta \mathcal{L}}{\delta (\partial_\mu \phi)}\delta_\alpha(\partial_\mu \phi) = \kappa = \partial_\mu (\frac{\delta \mathcal{L}}{\delta(\partial_\mu \phi)}\delta_\alpha\phi)
\end{equation}
which in terms of the conservation of the current reads explicitly
\begin{equation}
\label{modnoether}
    \partial_\mu (\frac{\delta \mathcal{L}}{\delta(\partial_\mu \phi)}\delta_\alpha\phi) = \frac{\delta\mathcal{L}}{\delta O} \, \delta_\alpha O
\end{equation}
Not only do we learn that the Noether current is no longer conserved, $\partial_\mu J^\mu \neq 0 $, but we also get an explicit expression for the non-conservation in terms of the transformation of the symmetry violating term in the Lagrangian. This will be a key ingredient in our field theory calculation below.

\section{Field theory computation}\label{sec:cft}
\subsection{Field theory computation in the Single-sided Scenario}
Start with the Lagrangian \eqref{singlesided}
\begin{equation}
\mathcal{L} = \mathcal{L}_0- h 
\left ( O_1 O_2 + O_1^* O_2^* \right ) 
\end{equation}
where we spelled out the ``$+c.c.$" explicitly to make sure we correctly account for factors of 2.
Under a $U(1)$ transformation that is a symmetry of the undeformed CFT Lagrangian ${\cal L}_0$ the full Lagrangian transforms as
\begin{equation}
 \delta\mathcal{L}= \frac{\partial \mathcal{L}}{\partial O_1} \delta O_1 + \frac{\partial \mathcal{L}}{\partial O_2} \delta O_2 +
 \frac{\partial \mathcal{L}}{\partial O_1^*} \delta O_1^* + \frac{\partial \mathcal{L}}{\partial O_2^*} \delta O_2^* 
 .  
\end{equation}
The operators $O_i$ having charge $q_i$ under the $U(1)$ global symmetry means their transformations are given by
\begin{equation}
\begin{aligned}\\&
O_i \rightarrow e^{i \alpha q_i} O_i = (1 + i\alpha q_i) O_i \\&
\delta O_i = i \alpha q_i O_i,
\quad \delta O_i^* = -i \alpha q_i O_i^*. \\&
\end{aligned}
\end{equation}
implying
\begin{equation}
     \delta\mathcal{L}= - i h \alpha (q_1 + q_2) \left ( O_2 O_1 - O_2^* O_1^* \right ) .
\end{equation}
The generalized Noether theorem of \eqref{modnoether} then gives
\begin{equation}
\label{nonconserved}
    \partial_\mu (J^\mu)  = -ih (q_1+q_2) (O_2 O_1 - O_2^* O_1^*)
\end{equation}
As in \cite{Aharony:2006hz}, the easiest way to process this information is to employ the operator-state map. The non-conservation equation \eqref{nonconserved} implies that at order $\mathcal{O}(h^2)$, the norm of the corresponding state $|\partial_{\mu} J^{\mu} \rangle$ is given by

\begin{eqnarray}
\nonumber
    \langle \partial_\mu J^\mu| \partial_\nu J^\nu \rangle &=&  h^2 (q_1+q_2)^2 \left ( \langle O_1 O_2 | O_1 O_2 \rangle 
    + \langle O_1^* O_2^* | O_1^* O_2^* \rangle  \right )
    \\ &=&
    2 h^2 (q_1+q_2)^2 \langle O_1 | O_1 \rangle \langle O_2| O_2\rangle = 2 h^2 (q_1+q_2)^2 N_1 N_2
    \label{fromnoether}
\end{eqnarray}
Here the norms of the states $|O_i \rangle$ are given by the corresponding 2-pt functions coefficients:
\begin{equation}
\langle O^*_i(x) O_i(0) \rangle = \frac{N_i}{x^{2 \Delta_i}} \, \quad \rightarrow \quad \langle O_i | O_i \rangle = N_i .
\end{equation}
The norm of $|O_i^* \rangle$ is identical to that of $|O_i \rangle$ and the two are orthogonal to each other.

There is an alternate way to calculate the norm of $| \partial_{\mu} J^{\mu} \rangle$. Under the operator state map the hermitian conjugate of the momentum operator $P_{\mu}=- i\partial_{\mu} $ is the special conformal generator $K_{\mu}$ so that 
\begin{equation}
    P_\mu |O\rangle \quad
    \leftrightarrow \quad \langle O| K^\mu .
\end{equation}
Furthermore, for primary operators the state $|O \rangle$ gets annihilated by $k_{\mu}$
\begin{equation}
    K_\mu |O\rangle 
    =0
\end{equation}
With this we can calculate the norm of $|\partial_{\mu} O \rangle$ and, more importantly for us, $|\partial_{\mu} J^{\mu} \rangle$. 
\begin{equation}
\begin{aligned}&
  | \partial_\mu O|^2 =   |P_\mu O|^2 = \langle O |K_\mu P^\mu| O\rangle = \langle O |\left[K_\mu, P^\mu \right]| O\rangle = -2 i (\delta_{\mu \nu} D-M_{\mu \nu})\\&
    |\partial_\mu J^\mu|^2 = \langle J_\nu |-2 i (\delta_{\mu \nu} D-M_{\mu \nu})|J^\mu\rangle
\end{aligned}
\end{equation}
The state $|J_{\mu} \rangle$ is an eigenstate of $-iD$ with eigenvalue $\Delta_J$, the dimension of the current operator. The action of $M_{\mu \nu}$ is that of an angular momentum generator on a vector. Last but not least, the norm of the state $|J_{\mu} \rangle$ is given by
\begin{equation}
\langle J_{\mu} | J_{\nu} \rangle = \delta_{\mu \nu} C_J
    \end{equation}
where the central charge $C_J$ is once more related to the coefficient of the 2-pt function of two current operators
\begin{equation}
    \langle J_{\mu}(x) J_{\nu}(0) \rangle = C_J \frac{\delta_{\mu \nu} - x_{\mu} x_{\nu}/x^2}{x^{2 \Delta_J}}.
    \label{currentwopoint}
\end{equation}
With this we have 
\begin{equation}
    \begin{aligned}&
      \langle J_\nu|-2 i (\delta_{\mu \nu} D)|J^\mu \rangle =2d \Delta_J d \\&   
      -2 \langle J_\nu |iM_{\mu \nu})|J^\mu\rangle = -2(d-1) C_J \delta^{\mu}_ {\mu} = -2 (d-1) d C_J .&
    \end{aligned}
\end{equation}
and collectively, 
\begin{equation}
\langle \partial_\mu J^\mu| \partial_\nu J^\nu \rangle=
    \langle J_\nu |-2 i (\delta_{\mu \nu} D-M_{\mu \nu})|J^\mu\rangle = 2d C_J (\Delta_J-d+1) 
    \label{confalg}
\end{equation}
Comparing \eqref{fromnoether} derived from the modified Noether's theorem with \eqref{confalg} that we got from using the conformal algebra, we find
\begin{equation}
    (\Delta_J - d+1) = \frac{h^2 (q_1+q_2)^2 N_1 N_2}{ d C_J} .
\end{equation}

The conformal dimension $\Delta_J$ of the current is related to the mass of the bulk photon by the standard AdS/CFT relation for a spin $p=1$ form \cite{Witten:1998qj}
\begin{equation}
    m^2 = (\Delta_J -p) (\Delta_J +p-d)
    =(\Delta_J -1) (\Delta_J +1-d)
\end{equation}
Writing
\begin{equation}
    (\Delta_J -1) = (\Delta_J -d+1)+ (d-2)
\end{equation}
we get 
\begin{eqnarray}
\nonumber
    m^2 &=& \left [ (\Delta_J -d+1)+ (d-2) \right ] (\Delta_J-d+1)\\
    &=& \nonumber
    \left[ \frac{h^2(q_1 +q_2)^2}{dC_J} N_1 N_2 \right]^2 + \frac{d-2}{d C_J} h^2(q_1 +q_2)^2 N_1 N_2\\
    &=& \frac{d-2}{d C_J} h^2(q_1 +q_2)^2 N_1 N_2 + {\cal O}(h^4) . 
    \label{massone}
\end{eqnarray}
\eqref{massone} constitutes our field theory prediction for the mass of the vector boson, obtained from calculating the anomalous dimension of the current operator due to the non-conservation induced by the double trace deformation in the single-sided scenario. Note, as a nice consistency check, that the mass vanishes in the special case $q_1=-q_2$. For this choice the double trace operator $O_1 O_2$ is actually neutral under the global $U(1)$ and so the symmetry is preserved. Reassuringly, this is exactly the case for which the mass vanishes.

\subsection{Field theory computation-Double-sided Scenario}

The calculation in the double-sided scenario runs very similar to the one in the single-sided scenario, so we'll mostly highlight the differences. This time we start with the Lagrangian from \eqref{doublesided}:
\begin{equation}
 \mathcal{L} = \mathcal{L}_1+\mathcal{L}_2 -h \left ( O_1 O_2 + O_1^* O_2^* \right ) 
\end{equation}
whose symmetry variation once more takes the form
\begin{equation}
 \delta\mathcal{L}= \frac{\partial \mathcal{L}}{\partial O_1} \delta_\alpha O_1 + \frac{\partial \mathcal{L}}{\partial O_2} \delta_\beta O_2 + c.c.
\end{equation}
In contrast to the single-sided case, this time each $O_i$ transforms under its own separate $U(1)$ symmetry:
\begin{equation}
\begin{aligned}\\&
O_1 \rightarrow e^{i \alpha q_1} O_1 = (1 + i\alpha q_1) O_1 \\&
\delta O_1 = i \alpha q_1 O_1\\& 
O_2 \rightarrow e^{i \beta q_2} O_2 = (1 + i\beta q_2) O_2 \\&
\delta O_2 = i \beta q_2 O_2\\&
\end{aligned}
\end{equation}
so that the variation of the Lagrangian under the two $U(1)$'s reads
\begin{equation}
\begin{aligned}\\&
     \delta_{\alpha,\beta} \mathcal{L}= \frac{\partial \mathcal{L}}{\partial O_1} \delta_{\alpha,\beta} O_1 + \frac{\partial \mathcal{L}}{\partial O_2} \delta_{\alpha,\beta} O_2
     + c.c.\\&
     \delta_{\alpha,\beta} \mathcal{L}= -h O_2 i \alpha q_1 O_1 -h O_2 i \beta q_2 O_2 + c.c.
\end{aligned}
\end{equation}
This allows us to derive the non-conservation of the two $U(1)$ currents written compactly as
\begin{equation}
\begin{aligned}
     \delta_{\alpha, \beta}\mathcal{L}= \partial_\mu (\frac{\delta \mathcal{L}}{\delta (\partial_\mu \phi)} \delta_\alpha \phi)+ \partial_\mu (\frac{\delta \mathcal{L}}{\delta (\partial_\mu \phi)} \delta_\beta \phi)  + c.c. = \partial_\mu(\alpha J_1^\mu)+\partial_\mu (\beta J_2^\mu)   
\end{aligned}
\end{equation}
and hence
\begin{equation}
   \alpha  \partial_\mu  J_1^\mu +  \beta \partial_\mu J_2^\mu =\partial_\mu J^\mu= -i h O_2 O_1 (\alpha q_1 +  \beta q_2) + c.c.
    \label{noncons}
\end{equation}
We can extract the two non-conservation laws by setting $\alpha=0$ or $\beta=0$ respectively.

While neither $J_1^{\mu}$ nor $J^\mu_{2}$ are conserved, we can find a linear combination of the two that is in fact conserved, corresponding to the fact that the symmetry breaking leaves one linear combination of the $U(1)$'s unbroken. To construct this conserved current, the right hand side of \eqref{noncons} must vanish.
This can be achieved by setting $\alpha = q_2 \epsilon$ and $\beta = -q_1 \epsilon$. The corresponding current is
\begin{equation}
    \tilde{J}^\mu = q_2 J^{\mu}_1 -q_1 J^{\mu}_2
\end{equation}
However, we are interested in the non-conserved current which will take the form of:
\begin{equation}
    J^{\mu} = a J^{\mu}_1 +b J^{\mu}_2 .
\end{equation}
We want to isolate the linear combination that is orthogonal to the conserved current $\tilde{J}^{\mu}$, which requires
\begin{equation}
    a q_2 C_{J1} -b q_1 C_{J2} = 0
    \label{constraint}
\end{equation}
where $C_{Ji}$ are the coefficients from the current 2-pt functions of $J^{\mu}_i$ as in \eqref{currentwopoint}.
This gives $b =a \frac{q_2 C_{J1}}{q_1 C_{J2}}$. Set $\gamma\equiv \frac{q_2 C_{J1}}{q_1 C_{J2}}$. We can then fix the overall normalization by setting $a=1$ which, using equation \eqref{noncons}, gives 
\begin{equation}
    \partial_\mu J^\mu=   -i h O_2 O_1 ( q_1 +  \gamma q_2) + c.c.
    \label{consoftildecurrent}
\end{equation}
as the divergence of the non-conserved current.
From this expression we can, as before, calculate the dimension $\Delta_J$ of the non-conserved current by determining the norm of $|\partial_{\mu} J^{\mu} \rangle$ in two different ways. The calculation follows the steps from the last subsection, the only difference being an overall prefactor. From Noether's law \eqref{consoftildecurrent} we find 
\begin{equation}
    \langle \partial_\mu J^\mu| \partial_\nu J^\nu \rangle = 2 h^2 ( q_1 +  \gamma q_2)^2 N_1 N_2
    \label{doubleradquant}
\end{equation}
and from the conformal algebra, we have:
\begin{equation}
    \langle \partial_\mu J^\mu |\partial _\nu J^\nu \rangle = \alpha^2 \langle J_1^\mu +\gamma J_2^\mu |-2 i (\delta_{\mu \nu} D-M_{\mu \nu})|J_1^\nu +\gamma J_2^\nu \rangle =  2d C_J (\Delta_J-d+1)
    \label{doubleconformal}
\end{equation}
where 
\begin{equation} C_J = C_{J1} + \gamma^2 C_{J2}.
\label{fieldtheorycj}
\end{equation}
Extracting the mass from comparing eq. \eqref{doubleconformal} with eq. \eqref{doubleradquant}, the mass of the vector field will be:
\begin{equation}
\begin{aligned}\\&
    m^2 = ((\Delta_J -d+1)+ (d-2))((\Delta_J-d+1))
    = \frac{d-2}{d C_J} h^2( q_1 +  \gamma q_2)^2 +{\cal O}(h^4) .\\&
    \label{masstwo}
\end{aligned}
\end{equation}

\section{Massive photon on the gravity side}\label{sec:bulk}
On the gravity side, we can calculate the same mass of the vector field from the scalar loop correction to the vector propagator. The basic strategy has been described in \cite{Porrati:2003sa} and explained in detail in \cite{Duff:2004wh}. We will compute the photon self-energy and then extract the long-distance behavior of the spin 0 piece and match it with the spin 0 massless Goldstone boson. As emphasized before, the boundary conditions play a crucial role. In the undeformed case, the asymptotic behavior of a massive field near the boundary of AdS goes as \eqref{asymp} with $\alpha_i(x)=0$. After adding in either the single- or double-sided deformation, the boundary conditions become \eqref{bc}:
\begin{equation}
    \alpha_1(x)=h(2\Delta_2-d)\beta_2^*(x),\qquad \alpha_2(x)=h(2\Delta_1-d)\beta_1^*(x).
\end{equation}

Since we restrict ourselves marginal deformations, $\Delta_1+\Delta_2 =d $, we can rotate the fields into each other such that they obey the boundary conditions of the undeformed theory \cite{Aharony:2005sh}. 
The new fields
\begin{equation}
\rho_1 = \frac{1}{\sqrt{1+\tilde{h}^2}} (\phi_1 + \tilde{h} \phi_2^*), \quad
\rho_2 = \frac{1}{\sqrt{1+\tilde{h}^2}} (- \tilde{h} \phi_1 + \phi_2^*)
\end{equation}
with $\tilde{h} = (2 \Delta_1 -d)h$ obey the standard $\alpha_{\rho_i}=0$ boundary conditions for a scalar dual to operators of dimension $\Delta_1$ and $\Delta_2=d-\Delta_1$ respectively.
If we write $G_{\Phi}^i$ for the undeformed propagators of a {\it real} scalar $\Phi$ dual to an operator of dimension $\Delta_i$, the correlators of the $\rho$ -fields hence read
\begin{equation}
    \langle \rho_1 \rho_1 \rangle =
    \langle \rho_2 \rho_2 \rangle = \langle \rho_1^* \rho_2 \rangle = \langle \rho_1 \rho_2 \rangle =0 , \quad \langle \rho_1 \rho_1^* \rangle =  G_{\Phi}^1, \quad
    \langle \rho_2 \rho_2^* \rangle =  G_{\Phi}^2 .
\end{equation}
Transforming back to the $\phi$ fields we find that the non-vanishing $\phi$ propagators can be summarized in a 2 x 2 matrix
\begin{equation}
\label{propmatrix}
    G_\phi^{ij}=\left(\begin{array}{cc}
        \langle \phi_1 \phi_1^* \rangle & \langle \phi_1 \phi_2 \rangle \\ \langle \phi_1 \phi_2 \rangle & \langle \phi_2 \phi_2^* \rangle
    \end{array}\right) =
    \frac{1}{(1+\tilde{h}^2)}\left(\begin{array}{cc}
        G_\Phi^1+\tilde{h}^2G_\Phi^2 & \tilde{h}(G_\Phi^1-G_\Phi^2) \\
        \tilde{h}(G_\Phi^1-G_\Phi^2) & G_\Phi^2+\tilde{h}^2G_\Phi^1
    \end{array}\right).
\end{equation}
All other 2-pt functions involving the $\phi_i$ fields vanish. Note in particular that while the diagonal terms have the standard structure of connecting a $\phi$ and a $\phi^*$ field, the off-diagonal terms propagate $\phi$ into $\phi$ and $\phi^*$ into $\phi^*$. Since exchanging $\phi$ with $\phi^*$ in the expression for the current, this gives an extra sign in the current 2-pt functions we are about to compute. 

The two canonically normalized scalar fields $\phi_i$ in the bulk each with action \eqref{complexaction} allow us to define two currents $j^1$ and $j^2$, where 
\begin{equation}
    j_\mu^i = iq_i (\phi_i^* \partial_\mu \phi_i - \partial_\mu \phi_i^* \phi_i),  \quad i=1,2. \end{equation}

The correlators $\langle j^ij^j\rangle$ are then given by $q_iq_j$ times derivatives acting on $\phi_i\phi_j$ and $\phi^*_i \phi_j$ correlators via Wick contractions using \eqref{propmatrix}. In both the single- and double-sided scenario what we need is the 2-pt function of a particular linear combination of these bulk currents, $j=A j^1+B j^2$  coupling to the gauge field that is being Higgsed. 
The correlator of this current of interest is then given by 
\begin{equation} 
\langle j_\mu j_{\nu'}\rangle=\langle A^2\, j_\mu^1j_{\nu'}^1+AB(\, j_\mu^1j_{\nu'}^2+j_\mu^2j_{\nu'}^1)+ B^2 \, j_\mu^2j_{\nu'}^2\rangle .
\label{bulkcorrelator}
\end{equation}

We can easily calculate the various correlators. The 1-1 diagonal term reads
\begin{equation}
\label{diagonal}
    \langle j_\mu^1j_{\nu'}^1\rangle=-2q_1^2 (\partial_\mu\langle \phi_1\phi_1^*\rangle\partial_{\nu'} \langle \phi_1\phi_1^*\rangle-\langle \phi_1\phi_1^*\rangle\partial_\mu\partial_{\nu'} \langle \phi_1\phi_1^*\rangle).
\end{equation}
and similar for the 2-2 diagonal term.

The off-diagonal terms are 
\begin{equation}
\label{offdiagonal}
    \langle j_\mu^1 j_{\nu'}^2\rangle=\langle j_\mu^2 j_{\nu'}^1\rangle=-q_1q_2(-\partial_\mu\langle \phi_1\phi_2\rangle\partial_{\nu'} \langle \phi_1^*\phi_2^*\rangle+\langle \phi_1\phi_2\rangle\partial_\mu\partial_{\nu'} \langle \phi_1^*\phi_2^*\rangle)+c.c.
\end{equation}
Note that the off-diagonal terms effectively picked up an extra minus sign due to the structure of \eqref{propmatrix}, where off-diagonal terms give non-vanishing propagators between the $\phi$ fields themselves, not their complex conjugates.

Each term current-correlator looks like a single bubble diagram with a single scalar and an effective propagator 
\begin{equation}
\label{effectiveprop}
G_{eff} = v G_{\Phi}^1 + w G_{\Phi}^2.
\end{equation}
To see the effect on the photon propagator, we need to work out the photon self-energy due to a bubble with effective propagator
\eqref{effectiveprop}.

Let us set up our conventions. We mostly follow the conventions in  \cite{Freedman:1998tz}, from which we also take the established results for the holographic 2-pt functions, including normalizations. The bulk photons have a kinetic term $-F^2/(4 g^2)$ in the single sided and $-F_i^2/(4 g_i^2)$ in the double sided case. This way no factors of the gauge coupling $g$ appear in the vertex coupling the scalar to the photon, and the vertex only contributes factors of $q_i$.

Since AdS is a maximally symmetric space, the photon self energy can be decomposed in terms of two bitensor structures \cite{DHoker:1999bve}: 
\begin{equation}
\label{selfenergy}
\Pi_{\mu \nu'}(Z) = \alpha(Z)g_{\mu \nu'} +\beta(Z)n_\mu n_{\nu'}.
\end{equation}
Here $Z= - \cosh(\mu)$ where $\mu$ is the geodesic distance between $x$ and $x'$, and $n_{\mu} = D_\mu \mu$ and $n_{\nu'} = D_{\nu'} \mu$ are unit tangent vectors to the geodesic at the points $x$ and $x'$. Properties of these quantities are summarized in table 1 of \cite{DHoker:1999bve}. To identify the mass we take the large geodesic distance limit corresponding to $Z = \frac{-1}{\epsilon}\to -\infty$ where we expand around $\epsilon \to 0$. We'll extract the vector mass later by comparing to the decaying behavior of the scalar propagator as in \cite{Porrati:2003sa,Duff:2004wh}.  
The massless photon propagator in AdS satisfies \cite{DHoker:1999bve}
\begin{equation}
    -(\Box+d)G_{\mu\nu'}=g_{\mu\nu'}\delta-\nabla_\mu\Box^{-1}\nabla_{\nu'},
\end{equation}
This term has the two tensorial structures with fixed relative coefficient. For the massive photon, there will be an extra term proportional to $n_\mu n_{\nu'}$ from the extra scalar polarization. As in \cite{Porrati:2003sa,Duff:2004wh} this extra scalar contribution has a different long distance fall-off than the one already present in the massless propagator. We can extract this extra scalar contribution by looking at the the photon self energy in the limit $Z \to -\infty$ and extracting the term proportional to derivatives of a massless scalar propagator.

For a single scalar loop with effective propagator \eqref{effectiveprop} and corresponding current correlators \eqref{diagonal} and \eqref{offdiagonal}, the coefficients $\alpha$ and $\beta$ in \eqref{selfenergy} evaluate to
\begin{eqnarray}
    \nonumber
\alpha(\epsilon)&=&vw \,\left ( \alpha_1\epsilon^{2\Delta_1+1}+\alpha_2\epsilon^{d+1}+\alpha_3\epsilon^{2\Delta_2+1} \right ) (1+ \mathcal{O}(\epsilon^2),\\
    \beta(\epsilon)&=& vw \, \left ( \alpha_1\epsilon^{2\Delta_1+1}+\alpha_2\epsilon^{d+1}+\alpha_3\epsilon^{2\Delta_2+1}  + \beta_1 \epsilon^d \right ) (1+ \mathcal{O}(\epsilon^2) ) ,
    \label{eq:vector}  
\end{eqnarray}
The coefficients $\alpha_{1,2,3}$ can easily be determined but play no role for us. The only important aspect is that they are the {\it same} in both $\alpha$ and $\beta$. Indeed, the difference between $\alpha$ and $\beta$ for a massless vector falls off faster than $\epsilon$ in $d \geq 2$ \cite{DHoker:1999bve}, so this is the expected structure of a massless vector. The extra scalar contribution is encoded in the remaining non-local piece:
\begin{equation}
    \begin{split}
    \beta_1 &=    2^{-d}\pi^{-1-d}(d-2\Delta_1)\Gamma(\Delta_1)\Gamma(\Delta_2)\sin\left(\pi(\Delta_1-\frac{d}{2})\right)\\
    &=\frac{\Gamma(\Delta_1)\Gamma(\Delta_2)}{2^{d-1}\pi^{d}\Gamma\left(\Delta_1-\frac{d}{2}\right)\Gamma\left(\Delta_2-\frac{d}{2}\right)}
    \end{split}
\end{equation}
To extract the contribution of this particular scalar loop to the vector mass, we compare the $\beta_1$ contribution with that of a scalar Goldstone boson eaten by the photon.  According to the AdS Higgs mechanism, we can write the contribution to the self-energy from a massless spin-0 Goldstone exchange as 
\begin{equation}
\nabla_\mu\nabla_{\nu'}G_\Phi^0=2^{-d}\epsilon^d\frac{\Gamma(d)}{\pi^{\frac{d}{2}}\Gamma\left(\frac{d}{2}\right)}(\epsilon g_{\mu\nu'}+dn_\mu n_{\nu'})
    \label{eq:goldstone}
\end{equation}
where $G_\Phi^0$ is the propagator of the massless scalar. This means the ratio is given by \cite{Porrati:2001db,Duff:2004wh}
\begin{equation}
   g^2\frac{\beta(\epsilon)}{\nabla_\mu\nabla_{\nu'}G_\Phi^0} = - 2g^2 vw \, \frac{\Gamma(\Delta_1)\Gamma(\Delta_2)\Gamma\left(\frac{d}{2}\right)}{d\pi^{\frac{d}{2}}\Gamma(d)\Gamma\left(\Delta_1-\frac{d}{2}\right)\Gamma\left(\Delta_2-\frac{d}{2}\right)} + \ldots
   \equiv v w M^2
   \label{bigM}
\end{equation}
where the $\ldots$ stand for the massless vector pieces in $\beta$ that are unimportant for the mass. The prefactor of $g^2$ is due to the fact in our non-perturbative normalization the photon propagator itself is proportional to $g^2$, so the self energy diagram gives a correction $g^4 \Pi$ to the leading order $g^2$ propagator.

To get the total mass of the vector induced by the scalar loops, we can simply add up the various contributions to the quantity in \eqref{bigM} from the correlators in \eqref{diagonal} and \eqref{offdiagonal}. For each $\langle \phi_i \phi_j \rangle$ or $\langle \phi_i^* \phi_j \rangle$ propagator the relevant values of $v$ and $w$ are encoded in the matrix of \eqref{propmatrix}. In particular, to leading order in $\tilde{h}$ the 1-1 correlator corresponds to $v=1$, $w=\tilde{h}^2$, the 2-2 correlator to $v=\tilde{h}^2$, $w=1$, whereas the off-diagonal correlators come with $v=-w=\tilde{h}$, and an extra overall minus sign as we explained above. To find the right linear combinations we have to finally commit to whether we study the single-sided or double-sided scenario. In the single-sided case the one bulk photon couples to both currents with equal strength, $A=B=1$ in \eqref{bulkcorrelator}. That means in this case
\begin{equation}
\label{singlebulk}
    m^2_{single} = (q_1 + q_2)^2 \tilde{h}^2 M^2
\end{equation}.

In the double sided case, we have two photons and need to single out the one dual to the non-conserved current. It couples to a linear combination of $j_1$ and $j_2$ that parallels the corresponding linear combination in the field theory, that is $A=a$ and $B=b$, with $a$ and $b$ constrained by the orthogonality requirement \eqref{constraint}. As in the field theory calculation, we choose $A=1$, implying $B=\gamma=\frac{q_2 C_{J1}}{q_1 C_{J2}}$ and hence yielding a mass
\begin{equation}
\label{doublebulk}
    m^2_{double} = ( q_1 + \gamma q_2)^2 \tilde{h}^2 M^2.
\end{equation}

Last but not least, to compare to the CFT calculation we need to plug in the standard gravity answers for the normalization constants $N_1$, $N_2$, and $C_J$ from \cite{Freedman:1998tz}. For the scalars we have 
\begin{equation}
\label{scalarnorm}
    N_1N_2=-  (\Delta_1-\Delta_2)^2\frac{\Gamma(\Delta_1)\Gamma(\Delta_2)}{\pi^d\Gamma\left(\Delta_1-\frac{d}{2}\right)\Gamma\left(\Delta_2-\frac{d}{2}\right)}.
\end{equation}
For the vector in the single-sided scenario there we only have a single coupling $g$, which is the one that directly appears in the expression for the mass \eqref{bigM}. Its relation to the normalization of the current 2-pt function in the field theory is given by
\begin{equation}
\label{vectornorm}
    C_J=\frac{(d-2)\Gamma(d)}{2\pi^{\frac{d}{2}}\Gamma\left(\frac{d}{2}\right)g^2}.
\end{equation}

In the double-sided, the photon dual to the non-conserved current has an effective coupling constant
\begin{equation}
    \frac{1}{g^2}=\frac{1}{g_1^2} + \left(\frac{q_2 C_{J1}}{q_1 C_{J2}}\right)^2 \frac{1}{g_2^2}.
\end{equation}
This is the coupling $g$ that appears in the expression \eqref{bigM} for the mass.
The $g_i$ are related to the field theory norms $C_{Ji}$ by the formula \eqref{vectornorm} with an extra $i$ subscript. Correspondingly we have for the effective $C_J$ from \eqref{fieldtheorycj}
\begin{equation}
\label{doublevectornorm}
    C_J=C_{J1} + \gamma^2 C_{J2} =\frac{(d-2)\Gamma(d)}{2\pi^{\frac{d}{2}}\Gamma\left(\frac{d}{2}\right) } \left ( \frac{1}{g_1^2} + \frac{\gamma^2}{g_2^2} \right ) = \frac{(d-2)\Gamma(d)}{2\pi^{\frac{d}{2}}\Gamma\left(\frac{d}{2}\right)g^2} 
\end{equation}
just as in the single-sided case.

Plugging \eqref{scalarnorm} and \eqref{vectornorm} into \eqref{massone} we find perfect agreement  with the bulk calculation \eqref{singlebulk} and \eqref{bigM}, and similar agreement is also obtained between \eqref{masstwo} and \eqref{doublebulk}.

\section*{Acknowledgements}

This work was supported in part by the U.S. Department of Energy under Grant No. DE-SC0022021 and a grant from the Simons Foundation (Grant 651440, AK). 

\bibliographystyle{JHEP}
\bibliography{refs.bib}

\end{document}